\documentclass[conference]{IEEEtran}
\IEEEoverridecommandlockouts
\usepackage{tabularx,booktabs}
\usepackage{multirow}
\usepackage{tabularx}
\usepackage{cite}
\usepackage{amsmath,amssymb,amsfonts}
\usepackage{algorithmic}
\usepackage{graphicx}
\usepackage{textcomp}
\usepackage{xcolor}

\usepackage{amsmath}
\usepackage{array}
\usepackage{multirow}
\usepackage{pifont}
\usepackage{caption}

\newcommand{\xmark}{\ding{55}} % Definir o comando \xmark

\def\BibTeX{{\rm B\kern-.05em{\sc i\kern-.025em b}\kern-.08em
    T\kern-.1667em\lower.7ex\hbox{E}\kern-.125emX}}
\begin{document}

% \title{Open Source Platforms for the 5G SA Core Network: Low-Cost Solution and Performance Evaluation}

% \title{A Brief Performance and Qualitative Evaluation of Open-Source 5G Core Platforms}

\title{Open-Source 5G Core Platforms: A Low-Cost Solution and Performance Evaluation}

% \author{\IEEEauthorblockN{1\textsuperscript{st} Maria Katarine Santana Barbosa}
% \IEEEauthorblockA{\textit{Centro de Infrmática (CIn)} \\
% \textit{Universidade Federal de Pernambuco (UFPE)}\\
% Recife, Brasil \\
% mksb@cin.ufpe.br}
% \and
% \IEEEauthorblockN{2\textsuperscript{nd} Marcelo V. B. da Silva}
% \IEEEauthorblockA{\textit{Centro de Infrmática (CIn)} \\
% \textit{Universidade Federal de Pernambuco (UFPE)}\\
% Recife, Brasil \\
% mvbs3@cin.ufpe.br}
% \and
% \IEEEauthorblockN{3\textsuperscript{rd} Ednelson L. L. Cavalcanti}
% \IEEEauthorblockA{\textit{Centro de Infrmática (CIn)} \\
% \textit{Universidade Federal de Pernambuco (UFPE)}\\
% Recife, Brasil \\
% ellc@cin.ufpe.br}
% \and
% \IEEEauthorblockN{4\textsuperscript{th}  Kelvin L. Dias}
% \IEEEauthorblockA{\textit{Centro de Infrmática (CIn)} \\
% \textit{Universidade Federal de Pernambuco (UFPE)}\\
% Recife, Brasil \\
% kld@cin.ufpe.br}
% }

\author{\IEEEauthorblockN{1\textsuperscript{st} Maria Barbosa}
\IEEEauthorblockA{\textit{Centro de Informática} \\
\textit{Universidade Federal} \\ \textit{de Pernambuco} (UFPE)\\
Recife, Brasil \\
mksb@cin.ufpe.br}
\and
\IEEEauthorblockN{2\textsuperscript{nd} Marcelo Silva}
\IEEEauthorblockA{\textit{Centro de Informática} \\
\textit{Universidade Federal} \\ \textit{de Pernambuco} (UFPE)\\
Recife, Brasil \\
mvbs3@cin.ufpe.br}
\and
\IEEEauthorblockN{3\textsuperscript{rd} Ednelson Cavalcanti}
\IEEEauthorblockA{\textit{Centro de Informática} \\
\textit{Universidade Federal} \\ \textit{de Pernambuco} (UFPE)\\
Recife, Brasil \\
ellc@cin.ufpe.br}
\and
\IEEEauthorblockN{4\textsuperscript{th}  Kelvin Dias}
\IEEEauthorblockA{\textit{Centro de Informática} \\
\textit{Universidade Federal} \\ \textit{de Pernambuco} (UFPE)\\
Recife, Brasil \\
kld@cin.ufpe.br}
}

\maketitle

\begin{abstract}
An essential component for the Fifth Generation of Mobile Networks deployments is the 5G Core (5GC), which bridges the 5G Radio Access Network (RAN) to the rest of the Internet. Some open-source platforms for the 5GC have emerged and been deployed in Common Off-the-Shelf (COTS)-based setups. Despite these open-source 5GC initiatives following the 3GPP specifications, they differ in implementing some features and their stages in the timeline of 3GPP releases. Besides that, they may yield different performance to metrics related to the data and control planes. This article reviews the major open-source 5GC platforms and evaluates their performance in a 5G Standalone (SA) COTS-based testbed. The results indicate that Open5GS provides the best latencies for control plane procedures, OpenAirInterface offers the highest data rates, and Free5GC has the lowest resource consumption.

\end{abstract}


\section{Introduction}
\label{sec:Introduction}

    % Contexto: Busca por redes privadas de baixo custo e o aumento na utilização de soluções open source.
    % Problema: Uma gama de stacks abertas para serem amplamente utilizadas. Qual selecionar? Como identificar se a stack tem aquilo que eu preciso? 
    % Motivação: Compreender e esclarecer o estado da arte e os resultados da avaliação de desempenho em relação às pilhas de software de código aberto adotadas na rede central 5G por meio de uma SLR.
    % Objetivo: Por meio de uma avaliação quantitativa e qualitativa das pilhas 5G de código aberto mais representativas, identificar as suas vantagens/limitações e destacando direções de pesquisa e desenvolvimento.
    % Contribuição: Avaliação de stacks open source, tanto qualitativamente quanto quantitativamente. O que mais? Identificar as lacunas no estado da arte e trazer mais pontos para cá. 

The 5G core network (5GC) plays a crucial role in the infrastructure of the Fifth Generation of Cellular Networks, connecting the 5G Radio Access Network (RAN) to other networks and the Internet \cite{5GSystem}. The 5GC is responsible for essential procedures such as authentication, access policies, session establishment, Quality of Service (QoS) provisioning, and charging. Differently from the core networks of previous generations, based on proprietary, closed, and monolithic infrastructures, the 5GC has embraced virtualization technology from its inception. Thus, data plane and control plane functions that were carried out by specialized physical equipment have migrated to virtualized Network Functions (NFs) \cite{5Gmicroservices}.

More recently, 5GC has been deployed based on open-source software solutions running on low-cost Commercial Off-The-Shelf (COTS) hardware \cite{MainStacks}. Particularly, the following open-source 5GC platforms stand out: Free5GC\footnote{https://free5gc.org/}, OpenAirInterface\footnote{https://openairinterface.org/oai-5g-core-network-project/} (OAI), Open5GS\footnote{https://Open5GS.org/}, and SD-Core\footnote{https://opennetworking.org/sd-core/}. By using those solutions on COTS-based setups, along with the appropriate spectrum licenses and permissions. Private 5G networks can be economically feasible to be deployed by governments to reduce the connectivity gap in underserved regions not assisted by the telecom operators, in industries to improve their processes and applications, and in research centers for experimentation, to cite a few possibilities \cite{IndustryAndAcademia}.

All aforementioned 5GC software stacks follow the 3GPP specifications but at different stages or with distinct implementation models. These platforms continue to evolve and are aligned with the new releases of 3GPP.  Therefore, it is essential to analyze and evaluate each of them to understand their benefits and limitations in more recent versions. The current state of the art has compared some of those open-source 5GC solutions qualitatively or quantitatively \cite{5GComparisonQuality}, \cite{5GCEvalQualQuant}, \cite{tutorial} and \cite{Lando2023}. However, to the authors’ knowledge, existing related work neither qualitatively analyzed all four platforms nor comparatively evaluated their performance in a single study based on an experimental setup.

This article first aims to provide a qualitative analysis of major open-source 5GC platforms, highlighting their main features, available resources, and limitations. Next, a prototype is built using general-purpose and programmable hardware to conduct a comparative performance evaluation among the targeted 5GC platforms. The testbed comprises Ettus Research B210 software-defined radio (SDR) boards for the 5G RAN and COTS servers to run the 5GC. Finally, the article highlights the results obtained by metrics evaluation related to the data/control planes and resource consumption.  

The next sections are structured as follows: Section \ref{sec:Geral5GC} presents concepts of the 5G Core. Section \ref{sec:Qualitarive} details the 5GC platforms. Section \ref{sec:RelatedWork} describes related work addressing quantitative and qualitative analysis of 5GC. The conception of a low-cost prototype for the 5G network is presented in Section \ref{sec:rede5G}. Section \ref{sec:Evaluation} presents the performance evaluation results. Finally, the conclusions and future work are presented in Section \ref{sec:conclusion}.

\section{5G Core Network Architecture}
\label{sec:Geral5GC}

\begin{figure}[ht!]
    \centering
    \includegraphics[width=0.28\textwidth]{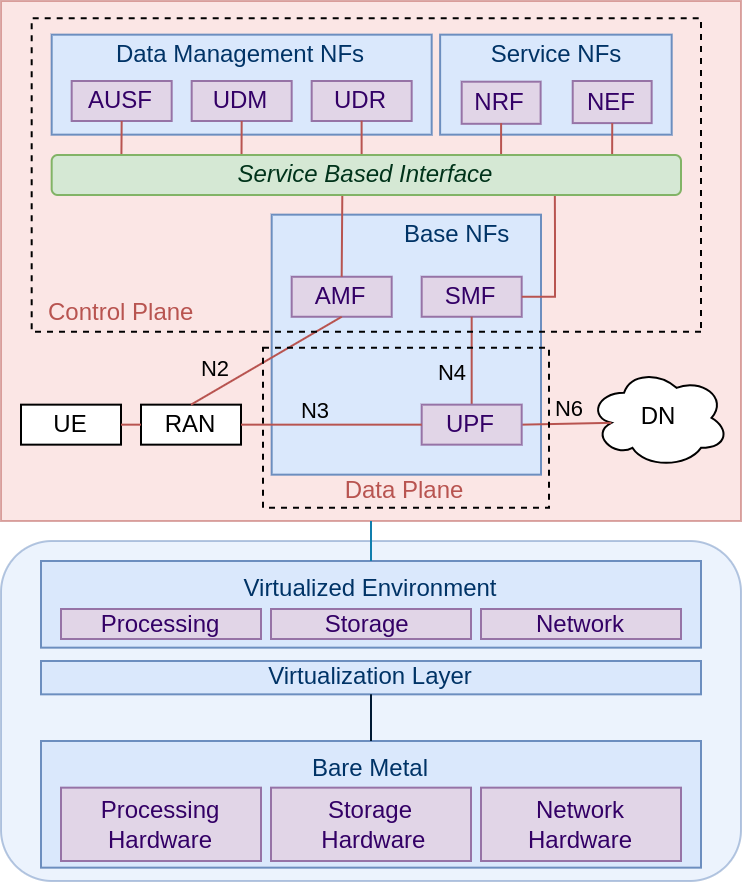}
    \caption{Overview of the 5G Core Network Architecture.}
    \label{fig:overview5GC}
\end{figure}

The 5GC is the cellular network component responsible for crucial functionalities for the system's operation and management. Figure \ref{fig:overview5GC} presents the major 5GC NFs responsible for management and signaling procedures. These NFs belong to the control plane (CP) or the data plane (UP). Moreover, Figure 1 also shows that 5GC NFs sit on top of a virtualized infrastructure.

In release 14, 3GPP introduced Control and User Plane Separation (CUPS) \cite{3GGPTSCups}, a critical element in the design of 5GC, as it allows for the independent scaling and distribution of NFs. This capability meets the minimum network requirements for applications such as URLLC, which demand low latency. CUPS enable UP NFs to operate at the network edge, closer to users, while CP NFs reside in the cloud \cite{5GCostEval}.

Unlike previous generations, the 5GC architecture emphasizes flexibility and adaptability, adopting a cloud-native approach. Each component has a Service-Based Interface (SBI), allowing each NF to act as a server and offer a service accessible through this interface. When multiple NFs interact in a network, providing and consuming services, a service-based architecture (SBA) emerges, a fundamental concept for the design of the 5GC.

Among the NFs of the CP, the Base NFs include the Access and Mobility Management Function (AMF) and the Session Management Function (SMF). Additionally, User Data Management NFs provide authentication and user identification services, including the Authentication Server Function (AUSF), Unified Data Management (UDM), and Unified Data Repository (UDR). Furthermore, some NFs do not directly relate to the user but facilitate the operation of other NFs, such as the Network Repository Function (NRF), which assists in service discovery, and the Network Exposure Function (NEF), which enables the 5GC to expose its services to third parties.

The UP in the 5GC comprises only the User Plane Function (UPF), which performs packet routing and forwarding, applies policies, reports traffic usage, and enforces QoS. The paradigm shift toward SBA in 5GC has enabled the adoption of virtualized infrastructure, mainly Kubernetes pods and containers. This approach has effectively reduced costs and streamlined deployment processes. By being hardware-independent, these solutions can operate on commercial servers or within cloud environments.

\subsection{5G Core and User Signaling Flow}
\label{sec:flowSig}

\begin{figure}[h!]
\centering
\includegraphics[width=0.4\textwidth]{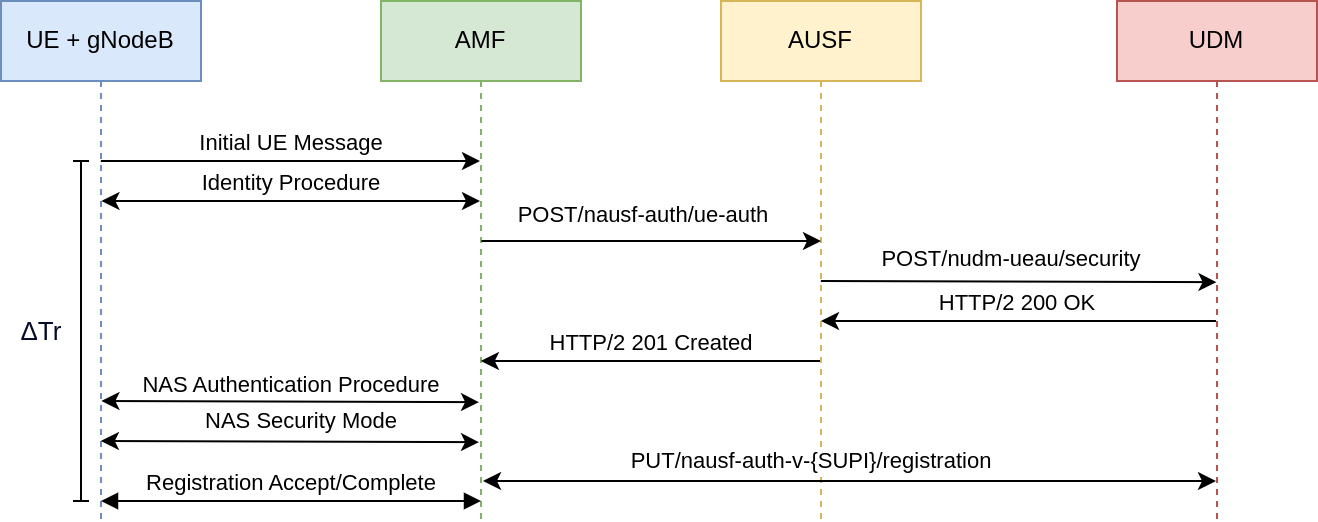}
\caption{Registration Signaling Flow.}
\label{fig:regflowsig}
\end{figure}

\begin{figure}[h!]
\centering
\includegraphics[width=0.4\textwidth]{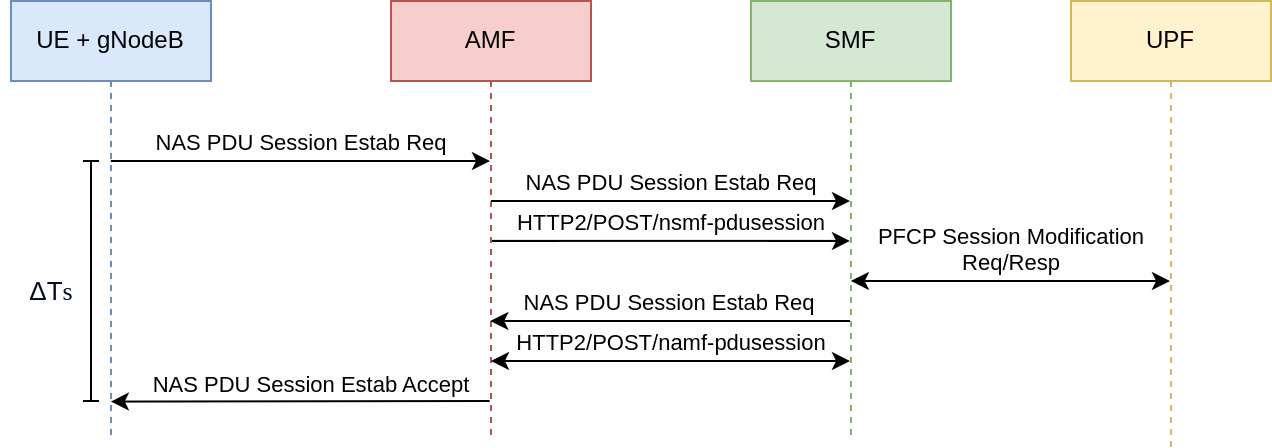}
\caption{PDU Session Establishment Signaling Flow.}
\label{fig:pduflowsig}
\end{figure}

Two procedures must be performed for the user equipment (UE) to establish 5G connectivity. The first is the registration process, which ensures that the UE establishes a connection with the network. The second procedure is the session establishment, which enables effective user data traffic. The signaling flows for these two procedures are presented in Figures \ref{fig:regflowsig} and \ref{fig:pduflowsig}. The Next Generation Node B (gNodeB) provides connectivity between the UE and the 5GC. In the figures, the UE and gNodeB are represented as a single block (UE + gNodeB), and the signaling messages between them are omitted since the RAN is not evaluated in this article.

The flows show two signaling: one between the UE and AMF, called Non-Access Stratum (NAS) signaling, and the between NFs that use SBI to communicate via the HTTP protocol. NF-to-NF communication can be a direct model where the NF consumer interacts directly with the NF producer, without NRF (Model A) or using a service discovery with NRF (Model B). On the other hand, in an indirect model, the communication is intermediate by a Service Communication Proxy (SCP). The consumer can communicate directly with the NRF (Model C) or via the SCP (Model D) \cite{3GGPTS}.

As illustrated in Figures \ref{fig:regflowsig} and \ref{fig:pduflowsig}, CP messages are exchanged one the 5GC NFs before data transmission, requiring time, $\Delta T_r$ for registration and $\Delta T_s$ for session establishment, and computational resources. Therefore, it is crucial that commercial hardware devices meet the minimum specifications needed to support this signaling.

\section{Open Source 5GC Platforms}
\label{sec:Qualitarive}

This section provides an overview of open-source 5GC platforms. Table \ref{tab:analise_qualitativa} highlights the main characteristics of each platform divided into two groups: the first outlines general specification, including 3GPP release, communication model between the NFs (NF Comm.), installation difficulty ranking (IDR), configuration difficulty ranking (CDR), programming language used for platform development, and virtualization infrastructure (VI). The second group presents the currently available features: Multi-access Edge Computing (MEC), Network Exposure Function (NEF), Data Plane Development Kit (DPDK), P4 programming, Web User Interface (WebUI), network monitoring (Net. Monitor), Voice over New Radio (VoNR), and support for non-3GPP devices (Non-3GPP Dev.).

\begin{table}[h!]
\centering
\caption{Comparison of Open Source 5GC Platforms.}
\begin{tabular}{ c c c c c }
\hline
\textbf{Parameter} & \textbf{Free5GC} & \textbf{OAI} & \textbf{Open5GS} & 
 \textbf{SD-Core} \\
\hline
\textbf{3GPP Rel.} & Rel-15 & Rel-16 & Rel-17 & Rel-16\\
\textbf{NF Comm.} & Model B & Model B & Model C &  Model B\\
\textbf{IDR} & 1 & 3 & 2 & 4 \\
\textbf{CDR} & 2 & 4 & 1 & 3 \\
\textbf{Language} & GO & C++ & C & GO \\
\multirow{2}{*}{\textbf{VI}} & Container & Container & Container & Pods\\
& and VM & & VM and Pods &  \\
\hline
\multicolumn{5}{c}{\textbf{Supported Features}} \\
\hline
\textbf{MEC} & \xmark & \checkmark & \xmark & \checkmark \\
\textbf{NEF} & \xmark & \checkmark & \xmark & \xmark \\
\textbf{DPDK} & \xmark & \checkmark & \checkmark & \checkmark\\
\textbf{P4} & \xmark & \xmark & \xmark & \checkmark \\
\textbf{WebUI} & \checkmark & \xmark & \checkmark & \checkmark \\
\textbf{Net. Monitor} &  \xmark &  \xmark & \checkmark & \checkmark \\
\textbf{VoNR} & \xmark & \xmark & \checkmark & \xmark\\
\textbf{Non-3GPP Dev.} & \checkmark & \xmark & \xmark & \xmark\\
\hline
\end{tabular}
\label{tab:analise_qualitativa}
\end{table}

Free5GC initially served as an academic test environment for 5G following release 15. However, this platform has gained commercial prominence because it supports non-3GPP devices such as Wi-Fi. The main ongoing developments are the SBI compliance with 3GPP release 17, support for Time-Sensitive Networking (TSN) and unauthenticated registration, and the addition of NFs such as NEF and Trusted Non-3GPP Gateway Function (TNGF) for secure Wi-Fi connections.

OAI platform provides projects for the RAN (gNodeB) and the 5GC in its repository. As for supported emerging technologies, OAI stands out by offering features such as MEC support enabling local application deployment. OAI also makes available an optional configuration aiming to optimize the performance of the UP by promoting high transfer rates for UPF with DPDK, a collection of libraries for fast packet processing whose base files are available on its repository using UPF-Vector Packet Processing (VPP). In addition, it provides an API for third-party access to network information through NEF. Planned functionalities include a dashboard to configure the network and support location services.

Open5GS offers a joint implementation of the 4G core (EPC) and 5G in Standalone (SA) and Non-Standalone (NSA) modes. It stands out among other platforms due to its support of Kamailio IP Multimedia Subsystem (IMS), making available VoNR services, and its compliance with 3GPP release 17. This platform offers integrated APIs for network monitoring using Prometheus\footnote{https://prometheus.io/} and Grafana\footnote{https://grafana.com/}, facilitating visualization of computational resource consumption and enabling automatic orchestration based on this data. Like OAI, Open5GS also has an optional configuration to deploy the UPF using DPDK. 

%Incluir o texto sobre o SD-Core
SD-Core has been developed by the Open Networking Foundation (ONF) as part of the Aether project, aiming to provide a cloud-based Connectivity-as-a-Service running on a Kubernetes Pod. The SD-Core provides a dual-mode solution, providing 4G and 5G (SA and NSA) connectivity. It is based on and enhances other ONF projects, the 4G Open Mobile Evolved Core (OMEC) and Free5GC. One of the main enhancements is employing the DPDK and P4 programming support, which allows for customized packet behavior definition, enabling the UPF to run in Edge clouds, supporting different user cases, and maintaining high performance. In addition, the SD-Core also has dashboards that display several aspects of Aether's runtime behavior and network monitoring through the Aether Management Platform (AMP).

% Nível de dificuldade na instalação e configuração da rede - parte 1{interface gráfica, tutoriais/suporte e arquiteturas disponiveis}
This paper defined a scale from 1 (easiest) to 4 (hardest) to assess the difficulty of installation and configuration. The availability of tutorials and information determines instruction level, clarity, and completeness (i.e., whether any information is missing) and whether the commands work as expected. In terms of installing basic NFs on VI, Free5GC offers a quick and easy installation with comprehensive and organized tutorials based on Docker Compose and Helm, followed by Open5Gs, which provides an installation based on Dockerfiles. OAI installation is based on docker and is relatively quick, but it presents challenges due to the tutorials because they are in a sub-repository (oai-cn5g-fed), which makes access difficult. SD-Core has structured, and documented installation tutorials based on entrypoints. However, some modifications to the installation files, such as changes to the host network interfaces and username, are required.

% Nível de dificuldade na instalação e configuração da rede - parte 2 {interface gráfica, tutoriais/suporte e arquiteturas disponiveis}
 
Regarding the 5G network configuration process, including QoS metric parameterization, network name, access points, and user subscription, Open5GS, Free5GC, and SD-Core provide a WebUI that simplifies the configuration process. OAI, on the other hand, offers a variety of architectures, from minimalist to more complex ones with multiple network slices. However, the user enrollment process in OAI is not structured, as it lacks a WebUI and requires direct modification of configuration files and databases.

%Custo Benefício
Finally, all these platforms enable the assembly of a low-cost 5G SA network and are widely applicable in various use cases. However, they still do not offer 5G full potential, such as support for the Network Data Analytics Function (NWDAF), which provides network automation and service management optimizations through data analytics and Machine Learning models.
\section{Related Work}
\label{sec:RelatedWork}

This section presents the state of the art in open-source 5G Core (5GC) evaluation from both qualitative and quantitative perspectives. Table \ref{tab:rel_work} highlights the main aspects of studies related to this research work.

\begin{table*}[ht]
        \caption{Summary of Related Work.}
    \centering

    \begin{tabular}{ | c | p{3.5cm} |c | p{2cm} | p{7.5cm} | }
     \hline
     \textbf{Paper} & \textbf{5GC Platform} & \textbf{RAN and UE} & \textbf{Type of Analysis} & \textbf{Aspects and Parameters Analyzed}\\
     \hline
     \cite{5GComparisonQuality} & Magma, Free5GC, and Open5GS & Not Applicable & Qualitative & Infrastructure, CUPS, documentation, license, community, maturity, and codebase.\\
     \hline
     \cite{5GCEvalQualQuant} & Free5GC, OAI, and Open5GS & Simulation & Qualitative and Quantitative & License, language, and infrastructure. Latency and hardware consumption.\\
     \hline
     \cite{tutorial} & Free5GC, OAI, and Open5GS & Simulation & Qualitative & Signaling compliance. \\
     \hline
     \cite{Lando2023} & Free5GC and Open5GS & Simulation & Quantitative & Throughput, latency, and packet loss. CPU and RAM consumption. \\
     \hline
     This Paper & Free5GC, Open5GS, OAI and SD-Core & Real & Qualitative and Quantitative & Language, infrastructure, 3GPP release, communication model of NFs, and supported resources. Evaluation of CP and UP.\\
     \hline
\end{tabular}
\label{tab:rel_work}
\end{table*}

A qualitative analysis among the Magma, Open5GS, and Free5GC platforms was conducted in \cite{5GComparisonQuality} within the context of the Plat5G-BR project, led by CPqD. The following topics were considered in the comparison among the platforms: deployment infrastructure, user documentation, community activity, code maturity, main code base, 5GC function management, and control and data plane separation. However, the analysis conducted by the authors does not provide details on the features supported by the platforms, the exposure of resources to third-party services or network services, such as MEC, and VoNR.

In \cite{5GCEvalQualQuant}, qualitative and quantitative analysis among the Free5GC, Open5GS, and Open Air Interface platforms is proposed. The parameters considered in this study for the qualitative part consist of license, language, and infrastructure used. Additionally, the authors propose a relevance index, ranking OAI first and Free5GC last. The quantitative analysis considered the following metrics:  CPU consumption and latency. The evaluation was carried out in a simulated RAN, utilizing from two to eight UEs. Despite a well-founded comparative analysis, the authors did not present the features supported by the platforms nor provide performance results for control plane metrics.  

The authors in the paper \cite{tutorial} conducted compliance and robustness testing of open-source platforms: Free5GC, OAI, and Open5GS. The compliance testing, based on 3GPP Release 16 standards, evaluated eleven procedures, revealing that all projects, except Open5GS in the case of Generic UE Configuration Update, meet the expectations. The results highlighted variations in the behavior of the projects, especially in scenarios such as Registration, Authentication, Security, SMF Selection, UPF Selection, and NAS Flow Validation, emphasizing differences in maturity levels. However, a quantitative analysis of both control and data plane metrics, as well as the computational resources consumed, is missing.

The study by \cite{Lando2023} demonstrates that Software-Defined Networks (SDNs) are a viable alternative for deployment in 5G networks. Network performance depends on the combination of the chosen platform and the hardware (bare metal) used. The article proposes using benchmarks to compare different combinations of platforms, Open5GS and Free5GC, and hardware in various 5G network scenarios. The metrics considered include data throughput, latency, packet loss, processor load, RAM usage, and execution time. The experiments were conducted using simulated UEs. According to the article, Free5GC achieved the best data throughput performance, while Open5GS obtained better stability during the registration of multiple devices. However, the study lacks practical, real-world experimentation of the evaluated scenarios.

The last line of Table \ref{tab:rel_work} presented the contribution of this paper. This study aims to compare the 5GC platforms through qualitative and quantitative evaluations using a real environment based on SDR for the 5G RAN and a general-purpose server for the 5GC. The qualitative evaluation, discussed in Section \ref{sec:Qualitarive}, was divided into two stages: the first stage involves a general assessment of the specifications of each platform, considering the license, 3GPP release, virtualization infrastructure, communication model between NFs, and the programming language used for development. The second evaluation checks the features supported by each platform. Additionally, previous studies employed only simulation environments; therefore, this proposal aims to set up a prototype and conduct a performance evaluation of CP and UP components in an experimental environment.
\section{A Low-Cost 5G Network Prototype}
\label{sec:rede5G}

\begin{figure}[ht!]
    \centering
    \includegraphics[width=0.35\textwidth]{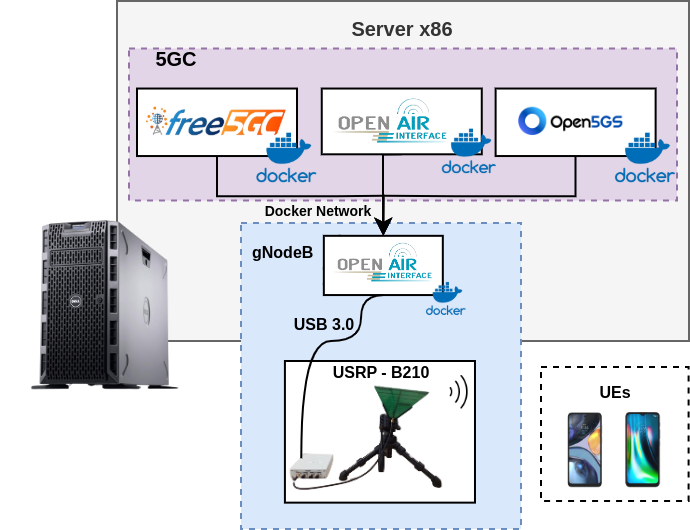}
    \caption{Architecture of the 5G SA network.}
    \label{fig:env}
\end{figure}

This section describes the hardware and software components used to design and deploy a 5G SA prototype network to evaluate the performance of the 5GC implemented by three open-source software platforms: Free5GC, Open5GS, and OAI. Since most 5GC reviewed in this paper are Docker container-based platforms, the SD-Core was excluded from the performance evaluation because it only supports Kubernetes pods. Figure \ref{fig:env} presents the proposed architecture for the 5G network testbed. Notice that all 5GC platforms implemented a common virtualization infrastructure. 

Table \ref{tab:confenv} presents the specifications for setting up the 5G network, such as those of the Docker Host server and the mobile device used. The testbed implements a Docker network on the server computer to establish communication among the 5GC (OAI, Free5GC, or Open5GS) and the gNodeB.    

The gNodeB comprises a joint OAI-based 5G NR implementation and SDR board. The server computer hosts a container running the components of the OAI stack for gNodeB and communicates with an Ettus B210 SDR board via a USB 3.0 port. The SDR conducts the baseband processing and generates the physical radio signal, which, together with an Ettus Log Periodic antenna, establishes 5G communication with real UEs for their connection and testing purposes. 

\begin{table}[h!]
    \centering
        \caption{Testbed Configuration.}
    \begin{tabular}{|c |p{3.5cm} |c|}
    \hline
     & \textbf{Component} &  \textbf{Specification} \\
    \hline
     \multirow{4}{4em}{\textbf{VI}} & CPU & Intel Xeon Gold 5215 \\ 
    & RAM & 96GB \\ 
    & \textit{Docker} & 24.0.5 \\
    & \textit{Docker Compose} & 2.20.2 \\
    & Operating System & Ubuntu 20.04.6 LTS \\
    \hline
     \multirow{7}{4em}{\textbf{RAN}} & Platform & Open Air Interface \\ 
    & SDR Board & Ettus B210 \\ 
    & Antenna & Ettus Log Periodic \\
    & UHD Version & 4.4.0.0\\
    & Band & n78 \\
    \hline
    \multirow{2}{4em}{\textbf{User}} & Model &  Motorola Edge 20\\
    & Sim Card & Sysmocom - S1J1. \\
    \hline
    \end{tabular}

    \label{tab:confenv}
\end{table}
\section{Performance Evaluation} 
\label{sec:Evaluation}

This section details the methodology for evaluating the performance of the 5GC platforms considered in this article, followed by the results and comparative analysis. The evaluation covers control plane metrics, including registration and session establishment times, and data plane metrics, such as end-to-end latency, throughput, and network behavior during video streaming requests. To establish a 95\% confidence interval, we collected 30 samples for each CP and UP tests and 50 samples for the resource consumption across all platforms.

\subsection{Control Plane Evaluation}

\begin{figure}[h!]
        \centering
        \includegraphics[width=0.36\textwidth]{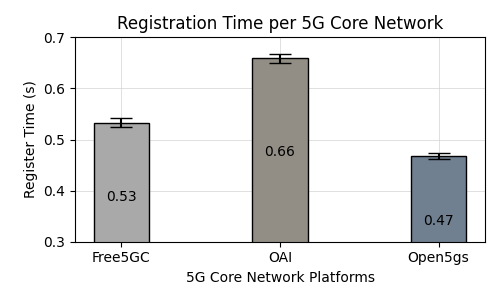}
        \caption{Registration Time Per 5GC.}
        \label{fig:eval_control_plane_1}
    \end{figure}

\begin{figure}[h!]
        \centering
        \includegraphics[width=0.36\textwidth]{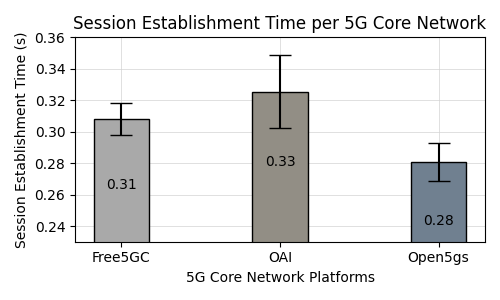}
        \caption{PDU Session Establishment Time Per 5GC.}
        \label{fig:eval_control_plane_2}
    \end{figure}

The CP metrics are UE registration time ($\Delta T_r$), which relies on AMF procedures, and session establishment time ($\Delta T_s$), involving the SMF. Both metrics rely on the signaling presented in Section \ref{sec:flowSig}. Figures \ref{fig:eval_control_plane_1} and \ref{fig:eval_control_plane_2} show the values of $\Delta T_r$ and $\Delta T_s$ for the three platforms. Open5GS has the smallest $\Delta T_r$ and $\Delta T_s$. Regarding $\Delta T_r$, Free5GC, and OAI are approximately 13\% and 40\% higher than Open5GS, respectively. For $\Delta T_s$, Free5GC, and OAI exhibit similar behaviors and continue to show values approximately 11\% and 17\% higher than those of the Open5GS platform, respectively.

\subsection{Data Plane Evaluation}
\label{sec:dataplaneEval}
To evaluate the end-to-end impacts of the platform on applications' QoS, this article measured and analyzed the following metrics: throughput, latency, and loading time (time taken for video download). The data plan evaluation occurred in two ways. First, focusing on the end-to-end flow capacity allowed between UE and UPF, the Iperf3 tool generated intensive data flows to congest the data path, impacting downloading and uploading scenarios. Ping was chosen for latency, as it sends small-sized packets, ensuring measurement is not affected by factors other than latency. The average values ($\bar{x}$) and the Standard Deviation ($s$) for each platform appear in Table \ref{tab:eval1}. According to the results shown in the table, OAI presented the best performance across all analyzed metrics, achieving a 50\% improvement in both download and upload compared to Free5GC, which had the poorest results and a 10\% latency reduction.

% \begin{table}[h!]
%   \caption{End-to-End Evaluation of 5GC Platforms.}
% \centering
%   \begin{tabular}{lSSSSSSSS}
%     \toprule
%     \multirow{2}{*}{Metrics} &
%       \multicolumn{2}{c}{Free5GC} &
%       \multicolumn{2}{c}{OAI} &
%       \multicolumn{2}{c}{Open5Gs} &
%       \multicolumn{2}{c}{SD-Core} \\
%       & {$\mu$} & {$\lambda$} & {$\mu$} & {$\lambda$} & {$\mu$} & {$\lambda$} & {$\mu$} & {$\lambda$}\\
%       \midrule
%     Download & 66,5 & 9,53 & 100 & 11,3 & 86,8 & 6,75 & - & - \\
%     Upload & 3,11  & 0,35 & 4,8 & 0,4 & 4,7 & 0,26 & - & -\\
%     Latency & 14.1  & 3,77 & 12,7 & 3,02 & 13,6 & 3,32 & - & - \\
%     \bottomrule
%   \end{tabular}

%   \label{tab:eval1}
% \end{table}

\begin{table}[!ht]
  \caption{End-to-End Evaluation of 5GC Platforms.}
\centering
  \begin{tabular}{lcccccc}
    \toprule
    \multirow{2}{*}{Platform} &
      \multicolumn{2}{c}{Download (Mbps)} &
      \multicolumn{2}{c}{Upload (Mbps)} &
      \multicolumn{2}{c}{Latency (ms)} \\
      & {$\bar{x}$} & {$s$} & {$\bar{x}$} & {$s$} & {$\bar{x}$} & {$s$}\\
      \midrule
    Free5GC & 66.5 & 9.53 & 3.11 & 0.35 & 14.14 & 3.77  \\
    OAI & 100.1  & 11.3 & 4.8 & 0.4 & 12.7 & 3.02 \\
    Open5Gs & 86.8  & 6.75 & 4.7 & 0.26 & 13.6 & 3.32 \\
    % SD-Core & -  & - & - & - & - & - \\
    
    \bottomrule
  \end{tabular}

  \label{tab:eval1}
\end{table}

The second evaluation analyzes throughput and loading time using a video streaming application. For these experiments, a Flask\footnote{https://flask.palletsprojects.com/en/3.0.x/}-based Python application providing video services and recording QoS metrics was created and deployed in an Ubuntu container. This container was connected only to the UPF through the N6 interface. On the user side, a Java-based Android application requests a video (Big Buck Bunny 60fps/4K) from the server, downloads it, and watches it on the phone. The application calculates loading time and network throughput and then sends these metrics to the server.

\begin{figure}[h!]
    \centering
    \includegraphics[width=0.36\textwidth]{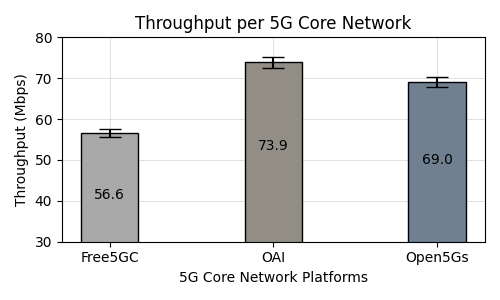}
    \caption{End-to-End Throughput Evaluation.}
    \label{fig:dataplaneeval_1}
\end{figure}

\begin{figure}[h!]
    \centering
    \includegraphics[width=0.36\textwidth]{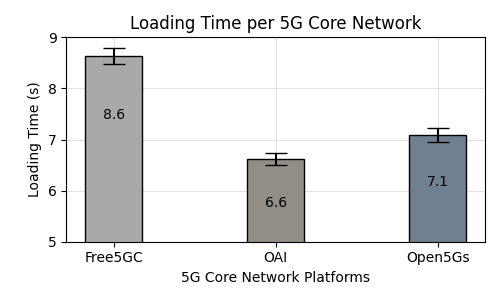}
    \caption{End-to-End Latency Evaluation.}
    \label{fig:dataplaneeval_2}
\end{figure}

Figures \ref{fig:dataplaneeval_1} and \ref{fig:dataplaneeval_2}  present the results obtained on the second evaluation. As can be seen, Open5GS and OAI platforms performed relatively closely together. Free5GC yielded the poorest results, with an approximately 23\% lower throughput and loading time 30\% longer than OAI. 

\subsection{Resource Consumption}

\begin{figure}[h!]
    \centering
    \includegraphics[width=0.35\textwidth]{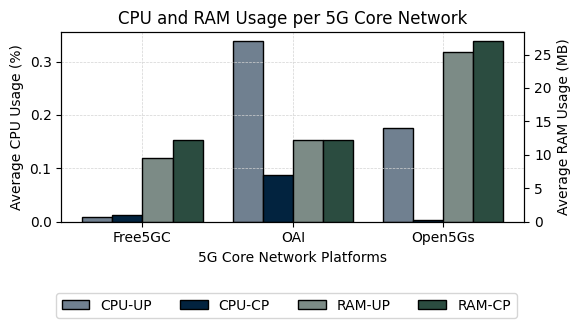}
    \caption{Average Resource Consumption.}
    \label{fig:resource}
\end{figure}

In addition to evaluating the control and data plane metrics, it is necessary to assess the hardware resources consumed for the signaling presented in Section \ref{sec:flowSig} (representing CP consumption) and the data traffic in Section \ref{sec:dataplaneEval} (representing UP consumption). Figure \ref{fig:resource} shows the average RAM and CPU consumption metrics for both UP and CP.

The figure shows that OAI exhibited the highest CPU consumption for both the CP and UP. On the other hand, Open5GS demonstrated 85\% and 45\% lower consumption for the CP and UP, respectively, when compared to OAI. Free5GC had the lowest CPU consumption, with a reduction of approximately 97\% relative to OAI for both the CP and UP. Regarding RAM consumption for the UP, Free5GC exhibited the lowest usage, followed by OAI, which consumed 28.97\% more RAM than Free5GC. Open5GS showed substantially higher RAM usage, approximately 165.42\% more than Free5GC. For the CP, OAI, and Free5GC had similar results, while Open5GS demonstrated RAM usage that was 121.61\% higher than OAI.
\section{Conclusion}
\label{sec:conclusion}

 % A rápida evolução das Redes Móveis impulsionada por plataformas de software de código aberto representa um marco significativo, resultado de colaborações entre acadêmicos, industriais e organizações de padronização. Embora tenhamos visto avanços notáveis com as pilhas de código aberto para o núcleo móvel 5G, ainda há desafios a serem superados. A necessidade de melhorar o desempenho para que se aproxime dos ambientes não virtualizados tradicionalmente empregados pelas operadoras, garantir suporte às funções exigidas pelo 3GPP e alcançar total compatibilidade com as infraestruturas nativas de nuvem e virtualizadas permanecem como objetivos cruciais. 

This paper provides an overview of the leading open-source 5GC platforms, followed by a preliminary performance analysis of these stacks. By highlighting the gaps and unique features of each 5GC solution, this paper may assist in choosing an appropriate open-source 5G core network for specific demands.The qualitative analysis highlighted critical features of the platforms and their supported capabilities. SD-Core offered the most supported emerging technologies, followed by Open5GS, OAI, and Free5GC. The quantitative analysis indicates that Open5GS achieved better latencies for control plane procedures. Open Air Interface gained advantages in the data plane, both under overload testing and for video streaming applications, and Free5GC has the lowest resource consumption. For future studies, it is essential to increase the number of users and address issues related to NF instantiation automation to optimize resource utilization. Additionally, future performance evaluations will encompass kubernetes-based infrastructure and the 5G SD-Core.

%visando compreender a correlação entre a quantidade de recursos utilizados e o número de usuários atendidos pela rede.

% 

% \section*{Acknowledgment}
% The authors thank the financial support provided for this work by FADE, the Foundation for Support of Development - UFPE, and the CIn/Motorola Project.

\section*{Acknowledgment}
This work was supported by the Motorola Mobility, National Council for Scientific and Technological Development (CNPq) - Research Productivity Fellowship (Grant No. 313083/2023-1) and Pernambuco Research Foundation (FACEPE) (Grant No. IBPG-0130-1.03/23).

\end{document}